\documentclass[pra,aps,onecolumn,epsfig]{revtex4}

\usepackage{graphicx}
\usepackage{rotating}
\usepackage{psfrag}

\begin{document}

\title{Checking C++ Programs for Dimensional Consistency}

\author{Ingo Josopait}

\affiliation{ Astrophysikalisches Institut Potsdam,
14482 Potsdam,
Germany}

\begin{abstract}
I will present my implementation 'n-units' of physical units into C++ programs.
It allows the compiler to check for dimensional consistency.

\end{abstract}
\maketitle

\newcommand{\etal}{\mbox{\it et.\ al.\ }}
\newcommand\iso[2]{$^{#2}$#1}
\newcommand\eqref[1]{(\ref{#1})}
\newcommand\myvec[1]{\vec{#1} \hskip 0.5mm}

\section{Introduction}

Computer simulations and other scientific programs often deal with physical quantities that have dimensional meanings, like length scales or time scales.
The internal representation of such quantities is done by floating point numbers.
The actual numbers have no direct meaning by themselves. Their meanings rely on the definition of the measuring units
(for example, the length '5 meters' could equally well be written as '500 centimeters' or '16.4 feet').

The addition, subtraction or comparison of two numbers of different dimensions, like time scales and length scales, is physically not meaningful and
can be regarded as an error. This follows from the principle of dimensional invariance, i.e. from the demand that the meaning of a formula should not depend on the choice of the
system of measuring units. Dimensional inconsistencies are a frequent source of errors in programs and much debugging time is usually 
spent to check a program for dimensional consistency.

However, the checking for dimensional consistency can be done automatically \cite{kennedy,allen}.
Implementations of units into programming languages like
python \cite{denis}
and C++ \cite{brown, dimnum}
exist.

I will present another implementation of units into the programming language C++. The source code is available at 
\verb@http://starburst.sourceforge.net/n-units/@.
The emphasis lies on computational speed. 
As in \cite{brown} and \cite{dimnum}, a check for dimensional consistency is done at compile time.

The main differences to these existing implementations are:
\begin{itemize}
\item Checking for dimensional consistency is designed to be deactivated for production runs, 
which results in better runtime performance (and reduced compile time).
\item Template definitions are simplified and the set of base units can easily be extended.
\item A function is provided that takes quantities to a fractional power.
\end{itemize}

\section{Dimensions, Units and Quantities}

Let me first give 3 definitions:

\paragraph{quantity}
A quantity is a property of some kind that can be quantified (e.g. the height of an object or its velocity).

\paragraph{dimension} A dimension specifies the type of a quantity (e.g. a length scale or a time scale). Only quantities
of the same dimension can be compared.

\paragraph{unit} A unit is a quantity that has been defined in order to measure other quantities and to be able to express them in terms of numbers.
Quantities can then be expressed as multiples of units. More than one unit can be defined per dimension. For example, cm, m and feet are all units
that represent length scales.

There are typically at least 3 independent dimensions used in a computer simulation:
\begin{itemize}
\item length scale
\item mass scale
\item time scale
\end{itemize}
This set can be extended. The Syst\`eme International d'unit\'es (SI) is based on 7 base units.
But also other dimensions like currencies, the amount of information (in bits or bytes), or the cosmological scale factor can be used.
More complex units (like energies or velocities) are derived from the base units.

\section{Checking for Dimensional Consistency}

A dimension is uniquely defined by the exponents of the base units. For example,
velocities (${\rm cm}^{1} {\rm s}^{-1}$) are composed of a length scale of exponent 1 and a time scale of exponent -1.
This can formally be expressed by vectors:
If we represent length scales by the vector (1,0,0), mass scales by (0,1,0) and time scales by (0,0,1),
velocities would be represented by the vector (1,0,-1).

For practical purposes it is sufficient to use a single integer number to represent the dimensional class.
This has two advantages:
\begin{itemize}
\item Template definitions are simplified.
\item Additional base units can be easily defined.
\end{itemize}

Quantities are represented by the following template class:
\begin{verbatim}
template <int n, class T=double> struct units
{
  T data;
  static units<n, T> construct(const T& a) 
  {
    // somewhat hidden constructor for explicit use
    units<n, T> r;
    r.data = a;
    return r;
  }

  ...
};
\end{verbatim}
The template parameter \verb@n@ specifies the dimension of the quantity, and \verb@T@ specifies the underlying floating point type.

\subsection{Base Units}

The base units are defined in the following way (as a convention, all units end with an underscore):\\
\begin{tabular}{ll}
\label{baseunits}
Length scale: & \verb+const units<1> m_ = units<1>::construct(1);+\\
Mass scale: & \verb+const units<10> g_ = units<10>::construct(1E-3);+ \\
Time scale: & \verb+const units<100> s_ = units<100>::construct(1);+ \\
...
\end{tabular}

This would define that the actual floating point representation follows the SI system (numbers are given in meters, kilograms and seconds) and
that the dimensions of length scale, mass scale and time scale are represented by the template parameters 1, 10 and 100, respectively.

Note that with the above definition the compiler cannot distinguish between, for instance, $\rm{cm}^{10}$ and $\rm{g}$. Since such large exponents of units are
rare, however, it is unlikely that this will be of practical importance.

\subsection{Basic Operations}
The following operations between quantities are allowed:
\begin{itemize}
\item Addition and subtraction are allowed between quantities of the same dimension.
\item Multiplication of two quantities of types \verb@units<@$m$\verb@>@ and \verb@units<@$n$\verb@>@ returns a quantity of type \verb@units<@$m+n$\verb@>@.
\item Dividing a quantity of type \verb@units<@$m$\verb@>@ by a quantity of type \verb@units<@$n$\verb@>@ returns a quantity of type \verb@units<@$m-n$\verb@>@.
\item Relational operators  ($==$, $<$, $>$) are allowed between quantities of the same dimension.
\end{itemize}

In the framework of C++ templates, this can be written as:

{
\small
\begin{verbatim}
template <int n, class T=double> struct units
{
  ...

  template <class B> units<n, typename addtype<T,B>::type> operator + (const units<n, B>& b) const; 
  template <class B> units<n, typename subtype<T,B>::type> operator - (const units<n, B>& b) const;
  template <int nb, class B> units<n+nb, typename multype<T,B>::type> operator * (const units<nb, B>& b) const;
  template <int nb, class B> units<n-nb, typename divtype<T,B>::type> operator / (const units<nb,B>& b) const;

  ...
};
\end{verbatim}
}

The classes \verb@addtype@, \verb@subtype@, \verb@multype@ and \verb@divtype@ are used to correctly determine the return
type of the underlying floating point number (so that, for instance, operations involving a \verb@float@ and a \verb@double@ always return a \verb@double@).

The dimensionless type \verb@units<0>@ is never used. Specializations of the above operators ensure that a bare floating point number (such as \verb@double@ or \verb@float@) is returned instead.

\subsection{Fractional Powers}
Taking fractional powers of quantities can be defined in the following way:
\begin{itemize}
\item Taking the square root of a quantity of type \verb@units<@$n$\verb@>@ returns a quantity of type \verb@units<@$n/2$\verb@>@.
\item More generally, taking a quantity of type \verb@units<@$n$\verb@>@ to the fractional power $p\over q$ returns a quantity of type \verb@units<@$n{p\over q}$\verb@>@.
\end{itemize}
For this purpose, the following template functions are provided:
\begin{verbatim}
template <int n, class T> units<n/2, T> sqrt(const units<n, T>& a);
template <int pa, int pb, int n, class T> units<(pa*n)/pb, T> pow(const units<n, T>& a);
\end{verbatim}
The expression $a^{p/q}$ can then be written as \verb@pow<p,q>(a)@, where \verb@p@ and \verb@q@ are constant integers.
Apart from the possibility to check for dimensional correctness, another advantage of using the above \verb@pow<>@ template function
is that the exponent $p\over q$ is known to the compiler. The template function \verb@pow<>@ can therefore attempt to use the 
standard functions \verb@sqrt(double)@ (which takes the square root) and \verb@cbrt(double)@ (which takes the cubic root)
in order to avoid the considerably slower function \verb@double pow(double, double)@.

\subsection{Data Types}

The dimension has to be specified for every quantity in the program.
The \verb@typeof@ extension of the gcc compiler is very useful for this. \verb@typeof(x)@ returns the type of 
the object \verb@x@. 
A velocity variable, for instance, can be defined by
\begin{verbatim}
typeof(cm_/s_) v  = 5*m_/s_;
\end{verbatim}
Because the units are defined to be constant, \verb@typeof@ expressions that involve only one unit should be written as a product to remove the constness.
A length \verb@h@ should therefore be defined as:
\begin{verbatim}
typeof(1*cm_) h = 2*km_;
\end{verbatim}

Unfortunately, the \verb@typeof@ keyword is not part of the ISO C++ standard.
However, it is possible to emulate the \verb@typeof@ extension \cite{gibbons} with the help of the \verb@sizeof@ keyword (which is part of the ISO C++ specification), at the cost of having to register every type (and dimension) to which \verb@typeof@ is applied.
To emulate \verb@typeof@, disable the \verb@HAVE_TYPEOF@ option in the header file.

The following small sample program illustrates the use of units in a program.
It calculates the time a slice of bread needs to fall from a table (of height 1 meter) to the floor.

\begin{verbatim}
#include <iostream>
#include "units.h"
using namespace std;

int main()
{
  typeof(1*cm_) height = 1*m_;
  typeof(cm_/s_/s_) g = 9.81*m_/s_/s_;
  typeof(1*s_) t = sqrt(2*height/g);
  cout << "free fall time=" << t/s_ << " seconds" << endl;
}
\end{verbatim}

Any violation of dimensional consistency would trigger a compiler error.

\section{Disabled Checking}
Dimensional checking can be disabled by the preprocessor option \verb@UNITCHECK@ in the header file. For performance reasons it is advisable to disable
it for production runs and to enable it only to check newly written code.
If \verb@UNITCHECK@ is disabled, the definitions of the base units (see section \ref{baseunits}) are replaced by constant floating point numbers:
\begin{verbatim}
const double m_ = 1;
const double g_ = 1E-3;
const double s_ = 1;
...
\end{verbatim}
The use of units will then have no negative influence on the runtime performance.

I would like to stress that even though the compile-time check of units in this implementation relies on the use of template classes,
units that are not checked for dimensional consistency can still be used in virtually any programming language, simply by defining the corresponding units as constant floating point numbers.

\section{Deriving Additional Units}

Once the set of base units is defined, other units can be derived from it, like for example:

\begin{verbatim}
const typeof(1*m_) cm_ = m_/100;                       // centimeter
const typeof(1*g_) kg_ = 1000*g_;                      // kilogram
const typeof(kg_*m_*m_/s_/s_) J_ = kg_*m_*m_/s_/s_;    // Joule
const typeof(m_/s_) c_ = 2.99792458e8 * m_ / s_;       // speed of light
\end{verbatim}
Given these definitions, units can be used directly in a program. One does not need to know the set of underlying measuring units
that is used to represent quantities. Even the combination of different units is possible. For example, the following expression is
perfectly valid:
\begin{verbatim}
typeof(1*cm_) height = 1*m_ + 75*cm_;
\end{verbatim}
The compiler will correctly add these two length scales. Since the system of base units is known at compile time, the compiler can optimize
this expression and perform the calculation during the compilation.

\section{Choosing the Base Type}

Sometimes the programmer wants to use a specific base type other than \verb@double@, like \verb@float@ or \verb@complex<>@. 
This can be accomplished either by explicitly using
\verb@cm_f@ (\verb@float@), \verb@cm_d@ (\verb@double@) or \verb@cm_ld@ (\verb@long double@) instead of \verb@cm_@ (which defaults
to \verb@double@) or by using the \verb@tof(..)@, \verb@tod(..)@, \verb@told(..)@ or \verb@to<>@ functions.
For instance, 
\begin{verbatim}
typeof(cm_f/s_f) v;
\end{verbatim}
and
\begin{verbatim}
typeof(tof(cm_/s_)) v;
\end{verbatim}
both define the velocity \verb@v@ to be of type \verb@float@.

\section{Output}

Since \verb@UNITCHECK@ should be deactivated for production runs, the compiler has no information about the dimensions of quantities.
Therefore, the programmer has to take care of meaningful output of quantities.
This can be done by dividing the quantity by the desired unit, for example:
\begin{verbatim}
void foo(typeof(m_/s_) v)
{
  cout << "v = " << v/(mile_/hour_) << " mph" << endl;
}
\end{verbatim}

\section{Conclusions}

I have presented an implementation of physical units into C++ programs. The code is checked for dimensional consistency at compile time.
The main advantages of this are:
\begin{itemize}
{
\item The programmer can use units directly in the code, without the need to know the system of base units.
\item Implementing complex formulae is simplified, because the programmer can be more relaxed about the dimensional correctness.
\item Dimensional correctness is guaranteed by the provided header files.
}
\end{itemize}

\bibliography{units}
\bibliographystyle{plain}

\end{document}